%% file: debye_mass-14-arx.tex
\documentclass[aps, 12pt, amsmath, amssymb, noshowpacs, preprint,prd]{revtex4-1}
\usepackage{amsmath}
\usepackage{amsfonts}
\usepackage{amssymb}
\usepackage{textcomp}
\usepackage[dvips]{graphicx}% Include figure files
\usepackage{bm}% bold math
\usepackage{color}

\usepackage{epsfig}
\usepackage{amsfonts,amsmath,amssymb}

\newcommand{\be}{\begin{equation}}
\newcommand{\ee}{\end{equation}}
\newcommand{\ba}{\begin{eqnarray}}
\newcommand{\ea}{\end{eqnarray}}
\newcommand{\ben}{\begin{enumerate}}
\newcommand{\een}{\end{enumerate}}

\newcommand{\ns}{\nonumber\\}

\newcommand{\dr}{\partial_{r}}

\newcommand{\lb}{\left(}
\newcommand{\rb}{\right)}

\newcommand{\ld}{\left.}

\newcommand{\rv}{\right|}
\newcommand{\lbr}{\left[}
\newcommand{\rbr}{\right]}
\newcommand{\ltr}{\left\langle}
\newcommand{\rtr}{\right\rangle}

\newcommand{\p}{\partial}

\newcommand{\la}{\langle}
\newcommand{\ra}{\rangle}

\newcommand{\rar}{\rightarrow}

\begin{document}
\title{Anomalous QCD Contribution to the Debye Screening in an External Field  via Holography.}

\author{A.~Gorsky}
\affiliation{ITEP, Moscow}
\author{P.~N.~Kopnin}
%\email{kopnin@itep.ru}
\affiliation{ITEP, Moscow}
\affiliation{MIPT, Moscow}
\author{A.~Krikun}
\affiliation{ITEP, Moscow} \affiliation{MIPT, Moscow}
\preprint{ITEP-TH-40/10}

\date{\today}
\begin{abstract}

In this paper we discuss the QCD contribution to the Abelian Debye
and magnetic screening masses in a deconfined QCD plasma at
finite temperature in the presence of an external magnetic field B.
We use a holographic AdS/QCD setup in an AdS Schwarzschield
black hole background and show that the electric screening mass has a form
similar to the one-loop result in QED. Moreover, we calculate the
corrections  due to the magnetic field to all orders of B and
demonstrate that in the case when magnetic field is large the Debye mass
grows linearly with B, while the magnetic screening mass
vanishes. The whole effect of the magnetic field turns out to stem
from the Chern--Simons action. We also discuss the zero
temperature case in the chiral perturbation theory.

\end{abstract}
%\pacs{11.25.Tq, 12.38.Lg, 12.39.Fe}

\maketitle

%\tableofcontents

%%%%%%%%%%%%%%%%%%%%%%%%%%%%%%%%%%%%%%%%%%%%%%%%%%%%%%%%%%%%%%%%%%%%%%%%%%%%%%%%%%%%%%%%%%%%%%%%%%%%%%%%%
%%%%%%%%%%%%%%%%%%%%%%%%%%%%%%%%%%%%%%%%%%%%%%%%%%%%%%%%%%%%%%%%%%%%%%%%%%%%%%%%%%%%%%%%%%%%%%%%%%%%%%%%%
%%%%%%%%%%%%%%%%%%%%%%%%%%%%%%%%%%%%%%%%%%%%%%%%%%%%%%%%%%%%%%%%%%%%%%%%%%%%%%%%%%%%%%%%%%%%%%%%%%%%%%%%%
%%%%%%%%%%%%%%%%%%%%%%%%%%%%%%%%%%%%%%%%%%%%%%%%%%%%%%%%%%%%%%%%%%%%%%%%%%%%%%%%%%%%%%%%%%%%%%%%%%%%%%%%%

\section{Introduction}

Debye screening is a well-known effect in quantum field theory. In
a hot plasma the static test charge is screened by real or virtual
charged particles of the medium. The screening potential has a
form $\dfrac{1}{r}\ e^{-r/l_D}$, where $l_D$ is the Debye
screening length. This potential emerges since photon
acquires the effective nonzero Debye mass $m_D = l_D^{-1}$
in certain external conditions.  A common method to
calculate the screening mass is to study the infrared behavior of
the polarization operator of the photon \cite{Kapusta}, \cite{Le
Bellac}. By definition, the electric screening mass is \be
m_D^2=e_q^2\Pi_{00}(\omega=0,\vec{k}^2=-m_D^2), \ee where \be
\Pi_{\mu\nu}(\omega,\vec{k})=\int d^4x\ \ltr
J_{\mu}(0)J_{\nu}(x)\rtr_{ret} e^{i\omega x_0 -
i\vec{k}\vec{x}}\label{el_mass} \ee is a polarization operator of
the photon corresponding to a retarded Green function and $e_q$ is
the charge with respect to the current $J$. A similar quantity
related to the $\Pi_{33}$ component of the polarization operator
is called the magnetic screening mass and reflects the screening
of the Lorentz force between two parallel currents: \be m^2_{D\
Mag}= e_q^2\Pi_{33}(\omega=0,\vec{k}=-m_D^2).\label{mag_mass} \ee

In QED the Debye mass at nonzero temperature is calculated in
perturbation theory using the ``hard thermal loop`` approximation.
The one loop result has been computed in \cite{Weldon} and equals
\be \label{one-loop} m_D^2 = \frac{e^2 T^2}{3}. \ee In
\cite{Blaizot} this quantity has been computed up to the $e^5$
order of perturbation theory. It is also interesting to study the
dependence of the Debye mass on an external magnetic field in
context of heavy ion collisions, where sufficiently large magnetic
fields \cite{Kharzeev,Skokov} may exist. The behavior of the
photon polarization operator and consequently the screening length
in nonzero external field was studied in \cite{Alexandre} by means
of the Schwinger proper time formalism \cite{Schwinger}. The
magnetic screening mass can be shown to vanish to all orders of
the perturbation theory \cite{Blaizot}.

The purpose of this paper is to compute the QCD effects on the
screening of the electromagnetic interaction. Clearly, the quark
loop enters the polarization operator of the photon, but one can
not limit oneself to one-loop approximation, because the coupling
constant of QCD is not small (at least unless the temperature is
too high). We study the QCD contributions to the screening masses
in two cases. At temperatures higher then the temperature of the
deconfinement phase transition, but low enough for the
non-perturbative treatment of QCD  ($1\mbox{ GeV}
\gtrsim T \gtrsim 200\mbox{ MeV} \approx T_c \sim \Lambda_{QCD}$),
we use the AdS/QCD model in the background of an AdS black hole
(BH) \cite{Maldacena,Witten,Son-Starinets,Herzog}. To study the
behavior of the Debye mass in an external magnetic field in the
case with confinement we use the chiral perturbation theory
approach at zero temperature \cite{ChPT} with the anomalous
Wess--Zumino--Witten term.

In the holographic calculation we are able to treat the external
magnetic field to all orders of the perturbation theory, thus
obtaining an exact analytical result for any values of $B$. It
turns out that the dependence on the external field is fully
driven by the Chern--Simons interaction in the action of the
AdS/QCD model. The Debye mass grows linearly with the magnetic
field in the strong field limit, thus coinciding with the behavior
found in \cite{Alexandre} via the mode analysis. At zero  magnetic
field the non-perturbative calculation gives the value of the
electric screening mass equal to the one-loop result in
perturbation theory (\ref{one-loop}). We also confirm
non-perturbatively the vanishing of the magnetic screening.

The paper is organized as follows. In Section \ref{section_b=0} we
give a brief introduction to the AdS/QCD methods and explore the
deconfinement region of the QCD phase diagram without the external
magnetic field. Section \ref{section_b!=0} is devoted to
holographical calculations in the same phase with an external
magnetic field. In Section \ref{section_chpt} we study the
confinement regime via the chiral perturbation theory. The
conclusion is given in Section \ref{section_conclusion}.

\section{Deconfined phase. $\mathbf{B=0}$\label{section_b=0}}

%Let us consider QCD at a temperature $T > T_c \approx
%\Lambda_{QCD}$. If the temperature is not extremely high, the
%system remains strongly coupled and one can use an AdS/QCD model
%to study it in a non-perturbative regime. In the deconfinement
%phase the geometry of bulk space is an AdS black hole
%\cite{Witten}.

In this section we set about calculating the Debye screening mass
$m_D$ in the absence of the magnetic field as defined in Eq.
(\ref{el_mass}), as well as the magnetic screening mass $m_{D\
Mag}$ which is defined analogously in Eq. (\ref{mag_mass}).

Since we are only interested in the QCD contributions to the mass
we therefore ignore all contributions of order of
$\alpha_{em}=e^2/4\pi$. One can easily notice that
$\Pi_{\mu\nu}(\omega=0,\vec{k}^2=-m_D^2)-\Pi_{\mu\nu}(\omega=0,\vec{k}=0)\propto
\alpha_{em}$, hence in what follows we shall study
$\Pi_{\mu\nu}(\omega=0,\vec{k}=0)$.

In order to obtain the screening masses, according to
(\ref{el_mass}, \ref{mag_mass}) one has to calculate a certain
two-point function. A holographic prescription for this
calculation \cite{Maldacena} states that one has to identify
five-dimensional fields dual to operators in question and assign
them boundary values equal to the sources of these operators (hence
the boundary conditions for the fields at the AdS boundary
$r=\infty$ are fixed). A classical five-dimensional action is then
identified with the logarithm of
the quantum field theory generating functional. Therefore, to
calculate a correlator via holography one has to vary the
classical action in the AdS with respect to the boundary values of
the relevant fields. In the calculation below we shall consider
two types of  correlation functions: the correlator of
temporal components of the electromagnetic currents in the case of
the Debye mass and the correlator of spatial components for the
magnetic screening mass. Hence we shall introduce the sources to
these currents in the corresponding cases. When investigating effects
at zero momentum it is quite handy to introduce a chemical
potential $\mu$ as a source of the temporal component of the
vector current. The source for the spatial component will be
denoted as $j$.

Let us consider the action of a holographic AdS/QCD model that
yields a dual description of QCD:

\be  S = S_{YM}[L]+S_{YM}[R]; \qquad S_{YM}[A] = -\
\frac{2}{8g^2_5}\int F\wedge\ast F .\nonumber \ee

It is a standard Abelian gauge sector of the AdS/QCD action, and
according to the AdS/CFT prescription the gauge fields are dual to
the QCD currents under consideration. For the sake of simplicity,
we are considering here a case with one quark flavor which
corresponds to the Abelian action, but a generalization to $N_f$
flavors is straightforward and does not lead to any qualitative
changes in our results. Note that the action has an additional
factor 2 as compared to that of the non-Abelian gauge fields
\cite{Erlich-Katz-Son-Stephanov}. This factor appears because the
OPE of the two-point correlation function of the current
$J_{\mu}=\bar{q}_f\gamma_{\mu}q_f$ of a particular quark flavor
$f$, which couples to the photon and is dual to the 5D gauge field
$A_\mu$ in our model, has the same factor 2 as compared to the OPE
of flavor-nonsinglet currents
$J^a_{\mu}=\sum\limits_{f,~f'}\bar{q}_f\gamma_{\mu}(t^a)_{ff'}
q_{f'}$ \cite{SUMrules}.The  $g_5$ is a 5D coupling constant and
is related to the number of colors
$\dfrac{R}{g_5^2}=\dfrac{N_c}{12\pi^2}$
\cite{Erlich-Katz-Son-Stephanov}. The action may also be rewritten
in terms of vector and axial gauge fields: $L=V+A,\ R=V-A$. \be
\label{actionYM} S = -\frac{1}{2g^2_5}\int dr\ d^4x\ \sqrt{-g} \lb
F^V_{MN}F^{V\ MN}+F^A_{MN}F^{A\ MN}\rb. \ee At temperatures under
consideration the quark condensate and all mesons are melted, and
thus the part of AdS/QCD action, responsible for a bifundamental
scalar (cf. \cite{Erlich-Katz-Son-Stephanov}),
 is absent.

The metric has the  form:
\ba ds^2 =
% g_{MN}dX^M dX^N =
\dfrac{r^2}{R^2}\lb
-f_{BH}(r)d^2t+d^2\vec{x} \rb +
\dfrac{R^2}{r^2}\dfrac{d^2r}{f_{BH}(r)},\
%\nonumber\\ &=&
%\frac{R^2}{z^2}\Big(-\tilde{f}_{BH}(z) dt^2 + dx_i^2 +
%\frac{1}{\tilde{f}_{BH}(z)} dz^2 \Big),\
\quad f_{BH}(r)=1-\frac{r_0^4}{r^4}.\label{metric}\ea
of an AdS Schwarzschield black hole \cite{Witten}. $R$ is
the AdS curvature radius, $r=\infty$ corresponds to the AdS
boundary, and the BH radius $r_0$ is related to the temperature of
the plasma: \be T=\frac{r_0}{\pi R^2} .\ee

Usually the presence of a nonzero chemical potential manifests
itself as a charge of a Reissner--Nordstr\"{o}m black hole, thus
altering the expression for the metric in (\ref{metric}). However,
in our calculations we are dealing only with two-point correlators
at zero chemical potential . The main terms of the
action itself are quadratic in $\mu$, while the account of the
black hole charge will yield the terms $\propto O(\mu^3),\
\mu\rightarrow 0,$ that cannot contribute to the two-point
functions at zero chemical potential. Therefore in what follows we
will keep the metric in the form of Eq. (\ref{metric}).

It has been pointed out in \cite{Son-Starinets, Herzog} that
calculations of retarded Green functions in AdS/CFT imply certain
boundary conditions at the horizon $r=r_0$: we have to make sure
that we leave only in-falling waves, which are solutions that are
regular at the horizon in the corresponding Eddington--Finkelstein
coordinates (see also \cite{Yee}). One can easily see that in the
case of a zero frequency this condition is equivalent to the
regularity of solution in standard AdS coordinates: \be A_i(r) \mbox{ and }
V_i(r) \mbox{ are regular at } r=r_0,\label{bc_regularity}\ee
where $i=1,2,3$. In addition , temporal components of the
gauge fields have to vanish at the horizon due to the fact that
$g_{00}(r=r_0)=0$: \be A_0(r_0)=V_0(r_0)=0\label{bc_zero}. \ee As
was pointed out earlier, the boundary condition at $r_0 \rar
\infty$ is determined by the source of the corresponding operator,
namely \be V_0(r=\infty)=\mu, \qquad V_3(r=\infty)=j,
\label{bc_infty} \ee \noindent .

Let us start with the simplest case of the
electric screening mass (\ref{el_mass}) in the absence of an
external field. Without the magnetic field the vector and the
axial gauge fields decouple. The only nonzero component of the
vector field in the present case is $V_0$,  hence the action (\ref{actionYM}) is reduced to :
\be S=\frac{N_c}{12 \pi^2 R^4}\int_{r_0}^{\infty} d^4x dr\ r^3 \lb
\partial_r V_0(r) \rb^2 .
\nonumber \ee Equation of motion for $V_0$ is quite trivial: \be
\dr(r^3\dr V_0)=0.\nonumber\ee
%The boundary conditions are the following: we demand that the
%solution is regular at the horizon $r=r_0$ and that $V_0$ vanishes
%there in order to allow for a nonsingular $V^0(r_0)$.
%\ba V_0(r=\infty)=\mu,\ V_0(r=r_0)=0. \label{vector_bc}\ea
The solution that takes into account the boundary condition at the
BH horizon (\ref{bc_zero}) and at the boundary (\ref{bc_infty})
reads as : \be V_0(r)=\mu\lb 1-\frac{r_0^2}{r^2} \rb,\nonumber\ee
and determines the value of the on-shell action: \be
S=V_{4D}\frac{\mu^2 N_c r_0^2}{6 \pi^2 R^4},\nonumber \ee where
$V_{4D}$ is the 4D volume. According to the holographic
prescription the correlator
$\Pi_{00}(\omega=0,\vec{k}=0)=\dfrac{1}{V_{4D}}\dfrac{\partial^2
S}{\partial \mu^2}$ yields the following value of the Debye mass
\be m_D^2=\frac{N_c}{3} e_q^2T^2.\label{el_mass_b0} \ee

Interestingly enough, the result of a non-perturbative QCD
calculation (\ref{el_mass_b0}) is similar to the leading term of
the QED perturbation series expression for the Debye mass
(\ref{one-loop}) in \cite{Weldon}.

Concerning the magnetic screening let us note that the equation of
motion for the spatial components of the vector field $V_i$ \be
\dr\lb r^3f_{BH}(r)\dr V_i(r) \rb=0\label{eqn_vector_spatial} \ee
allows only one solution which is regular at the horizon:
$V_i(r)\equiv const =V_i(\infty)$. Thus the action for the spatial
components $\propto \int dr\ r^3 f_{BH}(r)\lb\dr V_i(r)\rb$ is
zero, implying that the magnetic screening mass is zero: \be m_{D\
Mag}=0. \ee This result is in agreement with a statement that
$m_{D\ Mag}$ is zero to all orders of the perturbation theory (see
e.g. \cite{Blaizot}).

%%%%%%%%%%%%%%%%%%%%%%%%%%%%%%%%%%%%%%%%%%%%%%%%%%%%%%%%%%%%%%%%%%%%%%%%%%%%%%%%%%%%%%%%%%%%%%%
%%%%%%%%%%%%%%%%%%%%%%%%%%%%%%%%%%%%%%%%%%%%%%%%%%%%%%%%%%%%%%%%%%%%%%%%%%%%%%%%%%%%%%%%%%%%%%%
%%%%%%%%%%%%%%%%%%%%%%%%%%%%%%%%%%%%%%%%%%%%%%%%%%%%%%%%%%%%%%%%%%%%%%%%%%%%%%%%%%%%%%%%%%%%%%%
%%%%%%%%%%%%%%%%%%%%%%%%%%%%%%%%%%%%%%%%%%%%%%%%%%%%%%%%%%%%%%%%%%%%%%%%%%%%%%%%%%%%%%%%%%%%%%%
%%%%%%%%%%%%%%%%%%%%%%%%%%%%%%%%%%%%%%%%%%%%%%%%%%%%%%%%%%%%%%%%%%%%%%%%%%%%%%%%%%%%%%%%%%%%%%%
%%%%%%%%%%%%%%%%%%%%%%%%%%%%%%%%%%%%%%%%%%%%%%%%%%%%%%%%%%%%%%%%%%%%%%%%%%%%%%%%%%%%%%%%%%%%%%%
%%%%%%%%%%%%%%%%%%%%%%%%%%%%%%%%%%%%%%%%%%%%%%%%%%%%%%%%%%%%%%%%%%%%%%%%%%%%%%%%%%%%%%%%%%%%%%%

\section{Deconfined phase. $\mathbf{B\neq 0}$ \label{section_b!=0}}

\subsection{The action}

In this section we shall introduce the magnetic field by means of
the Chern--Simons (CS) action, see \cite{Yee,Chern-Simons}. The
full action of the model is now a sum of Yang-Mills
(\ref{actionYM}) and Chern--Simons terms. \ba S &=&
S_{YM}[L]+S_{YM}[R]+S_{CS}[L]-S_{CS}[R]\label{actionfull} \\
S_{CS}[A] &=& -\frac{N_c}{24\pi^2}\int A\wedge F\wedge F
-\frac{1}{2}A\wedge A\wedge A\wedge F +\frac{1}{10}A\wedge A\wedge
A\wedge A\wedge A \ns &=& -\frac{N_c}{24\pi^2}\int dz\ d^4x\
\epsilon^{MNPQR}A_M F_{NP} F_{QR}.\label{actionCS}\ea In the
Abelian case only the cubic term in the CS action is relevant. In
terms of the vector and axial fields $L=V+A,\ R=V-A$ it assumes
the form: \ba S_{CS} &=& \frac{-N_c}{4\pi^2}\int dr\ d^4x\
\epsilon^{MNPQR}A_M F^V_{NP} F^V_{QR} + \frac{-N_c}{12\pi^2}\int
dr\ d^4x\ \epsilon^{MNPQR}A_M F^A_{NP} F^A_{QR} \nonumber\\ &+&
\ld\frac{-N_c}{6\pi^2}\int d^4x\
\epsilon^{\mu\nu\lambda\rho}A_{\mu}V_{\nu}F^V_{\lambda\rho}\rv^{r=\infty}_{r=r_0}.\label{actionCS2}
\ea The Chern--Simons term  gives rise to the
interaction with the external magnetic field.
and $F^V_{12}(r=\infty)$ is associated with the magnetic field
multiplied by  the electric charge, $e_q B$.

There are two ways to obtain the expressions for the screening
masses in this setting. The first one is to treat the problem
perturbatively ,
considering Feynman diagrams that contain various numbers of legs
corresponding to the external magnetic field, which is carried
out in Subsection \ref{section_diagrams}. This consideration in
its turn motivates a non-perturbative diagonalization of the
action in the external field, which is performed in Subsection
\ref{section_eom}.

%%%%%%%%%%%%%%%%%%%%%%%%%%%%%%%%%%%%%%%%%%%%%%%%%%%%%%%%%%%%%%%%%%%%%%%%%%%%%%%%%%%%%%%%%%%%%%%
%%%%%%%%%%%%%%%%%%%%%%%%%%%%%%%%%%%%%%%%%%%%%%%%%%%%%%%%%%%%%%%%%%%%%%%%%%%%%%%%%%%%%%%%%%%%%%%
%%%%%%%%%%%%%%%%%%%%%%%%%%%%%%%%%%%%%%%%%%%%%%%%%%%%%%%%%%%%%%%%%%%%%%%%%%%%%%%%%%%%%%%%%%%%%%%
%%%%%%%%%%%%%%%%%%%%%%%%%%%%%%%%%%%%%%%%%%%%%%%%%%%%%%%%%%%%%%%%%%%%%%%%%%%%%%%%%%%%%%%%%%%%%%%
%%%%%%%%%%%%%%%%%%%%%%%%%%%%%%%%%%%%%%%%%%%%%%%%%%%%%%%%%%%%%%%%%%%%%%%%%%%%%%%%%%%%%%%%%%%%%%%
%%%%%%%%%%%%%%%%%%%%%%%%%%%%%%%%%%%%%%%%%%%%%%%%%%%%%%%%%%%%%%%%%%%%%%%%%%%%%%%%%%%%%%%%%%%%%%%
%%%%%%%%%%%%%%%%%%%%%%%%%%%%%%%%%%%%%%%%%%%%%%%%%%%%%%%%%%%%%%%%%%%%%%%%%%%%%%%%%%%%%%%%%%%%%%%

\subsection{Diagrams \label{section_diagrams}}
From the action (\ref{actionfull}) one gets equations of motion
for spatial and temporal components of the vector field in the
infrared limit ($\omega = 0, q^2 = 0$):
\begin{align}
-\p_r (r^3 \p_r V_0 (r)) &= (AV)\mbox{ interactions};\nonumber \\
-\p_r (r^3 f_{BH} (r) \p_r V_i (r)) &= (AV)\mbox{
interactions}.\nonumber
\end{align}
The bulk-to-boundary propagators $v_0$ and $v_i$ are solutions to
these equations without the interaction terms, subject to boundary
conditions (\ref{bc_zero}),(\ref{bc_regularity}),
(\ref{bc_infty}). (A more detailed study of the perturbation
theory in question may be found in \cite{Krikun-4point}.) Two
branches of the solution to the spatial equation are $v_i=1$ (see
Eq. (\ref{eqn_vector_spatial})) and
$v_i=\log\left(\frac{r^2-r_0^2}{r^2+r_0^2}\right)$. The latter
diverges at the horizon ($r=r_0$), so we must omit it. Hence we
end up with a trivial spatial bulk-to-boundary propagator:
\begin{equation}
v_i(r)|_{q,\omega=0} = 1.\label{spatial_propagator}
\end{equation}

\begin{figure*}[h!]
\flushleft{$a)$}
\\
\input{diagram1a.pstex_t}
\\
\flushleft{$b)$} 
\\
\input{diagram1b.pstex_t}

%\center{\includegraphics{figure1.eps}}
\caption{Tree-level diagrams, corresponding to calculations of the
vector current correlator in the external field $F_{12}$:
$\left.a\right)$ temporal components $\la J_0, J_0 \ra$ , $\left.b\right)$ spatial components $\la J_3, J_3
\ra$
.}\label{fig_diag}
\end{figure*}
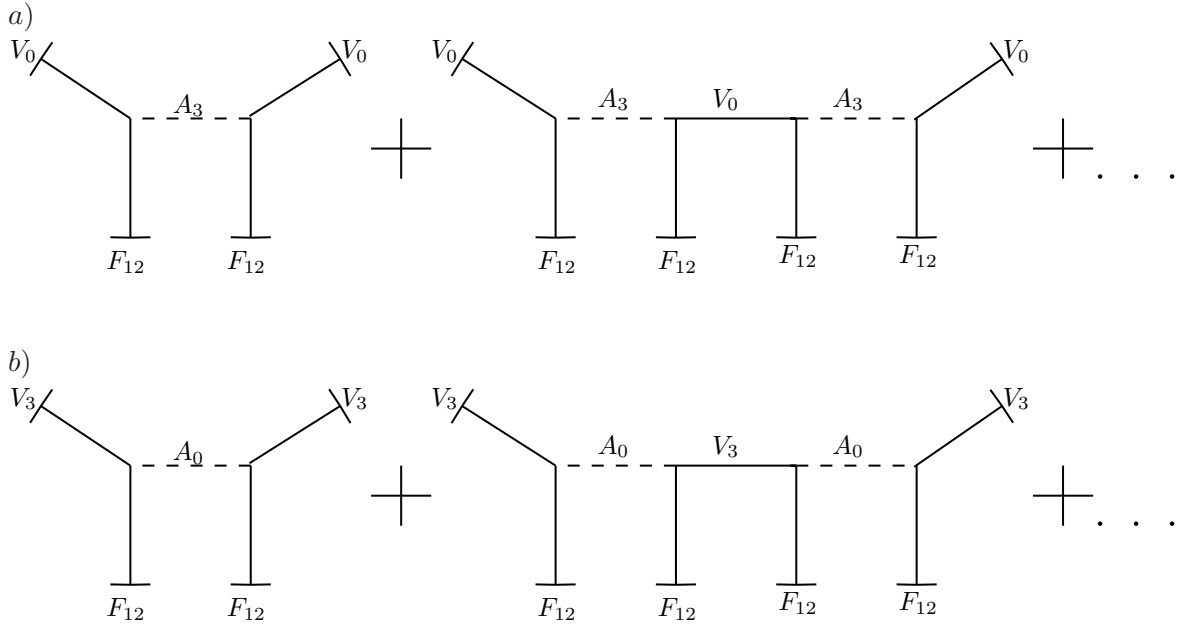

If we consider the Chern--Simons action in Eq. (\ref{actionCS2}),
we observe that both the terms involving three axial fields and
the term with one axial and two dynamical vector fields can
contribute to the two-point correlators. Moreover we  need to take
into account only the first term with two dynamical
axial and vectors fields and a vector field which
stems from the external magnetic field propagating into the bulk.
We also note that even in the case of a non-Abelian action
(involving $N_f$ quark flavors) vertices from the non-Abelian part
of the Yang-Mills action do not contribute to the correlator under
consideration (see \cite{Krikun-4point}).  We can depict the $\la
J_0 J_0 \ra$ and $\la J_3 J_3 \ra$ correlation functions that
determine the electric and the magnetic screening masses as a sum
of diagrams (Fig. \ref{fig_diag}) that include the aforementioned
vertex from the Chern--Simons term (while the correlators $\la J_1
J_1 \ra = \la J_2 J_2 \ra$ are discussed further below).
From the action (\ref{actionCS2}) we read out the vertex functional
\cite{Krikun-4point}:
\begin{equation}
\label{A_vertex}
\mathbb{A}_{\alpha \beta \gamma}=\delta^4(q_1 {+} q_2 {+}
q_3) \  \epsilon^{\alpha \beta \gamma
\sigma} \frac{r^2}{R^2}(\p_r^2 q_\sigma^1 - \p_r^1 q_{\sigma}^2).
\end{equation}

Due to the epsilon symbol all interacting fields must have
different Lorentz indices, hence there can only be one temporal
component at each vertex. As the spatial bulk-to-boundary
propagator is trivial, it can not acted on by $\p_r$, so the
only way to get a nonzero result is to act by $\p_r$ on the
temporal component. Furthermore, the momenta of all incoming
photons are zero, hence in a tree-level diagram all momenta should
be zero. Therefore the only way to place a spatial derivative is
to act with it on the external field and obtain the dual field
strength tensor $\tilde{F}^{\alpha \beta} = \epsilon^{\alpha \beta
\gamma \sigma} q_\sigma V_\gamma$. Ultimately we find that the
triple vertex boils down to a mixing term between the spatial
component of the axial field and the temporal component of the
vector field (Fig. \ref{fig_diag}a), or vice versa (Fig.
\ref{fig_diag}b). If we choose the external magnetic field to be
$(0,0,B)$, we get an effective interaction term : \ba
S_{CS} = &-&\frac{N_c e_qB}{2\pi^2}\int d^4x dr\ A_3(r)\dr V_0(r)
+ \ld\frac{N_ce_qB}{6\pi^2}\int d^4x\ A_3
V_0\rv^{r=\infty}_{r=r_0}\nonumber\\
&+&\frac{N_c e_qB}{2\pi^2}\int d^4x dr\ V_3(r)\dr A_0(r) -
\ld\frac{N_ce_qB}{6\pi^2}\int d^4x\ V_3 A_0\rv^{r=\infty}_{r=r_0},
\label{action_mixing} \ea where the first line is relevant for
the $\la J_0 J_0 \ra$ correlator --- see Subsection
\ref{section_eom_el} while the second one --- the $\la J_3 J_3 \ra$
correlator --- see Subsection \ref{section_eom_mag}.

A simple consideration of the diagrams demonstrates that the
quantities $\Pi_{11}$ and $\Pi_{22}$ are identically zero due to
the epsilon symbol in the vertex (\ref{A_vertex}). Note however,
that these quantities vanish only in the leading order of large
$N_c$ expansion. To find the $\frac{1}{N_c}$ corrections to this
result, one should consider the diagrams with dilaton and graviton
exchange in the bulk, which can produce nonzero input to
$\Pi_{11}$ and $\Pi_{22}$ (see \cite{Krikun-4point}). The
consistent treatment of $\frac{1}{N_c}$ corrections involves
taking account of higher orders of string perturbation theory,
which is out of the scope of this paper. Nevertheless, we can
state the result
\begin{equation}
\label{O(1)}
\Pi_{11}, \Pi_{22} = O(1),
\end{equation}
while $\Pi_{00} = O(N_c)$.

%%%%%%%%%%%%%%%%%%%%%%%%%%%%%%%%%%%%%%%%%%%%%%%%%%%%%%%%%%%%%%%%%%%%%%%%%%%%%%%%%%%%%%%%%%%%%%%
%%%%%%%%%%%%%%%%%%%%%%%%%%%%%%%%%%%%%%%%%%%%%%%%%%%%%%%%%%%%%%%%%%%%%%%%%%%%%%%%%%%%%%%%%%%%%%%
%%%%%%%%%%%%%%%%%%%%%%%%%%%%%%%%%%%%%%%%%%%%%%%%%%%%%%%%%%%%%%%%%%%%%%%%%%%%%%%%%%%%%%%%%%%%%%%
%%%%%%%%%%%%%%%%%%%%%%%%%%%%%%%%%%%%%%%%%%%%%%%%%%%%%%%%%%%%%%%%%%%%%%%%%%%%%%%%%%%%%%%%%%%%%%%
%%%%%%%%%%%%%%%%%%%%%%%%%%%%%%%%%%%%%%%%%%%%%%%%%%%%%%%%%%%%%%%%%%%%%%%%%%%%%%%%%%%%%%%%%%%%%%%
%%%%%%%%%%%%%%%%%%%%%%%%%%%%%%%%%%%%%%%%%%%%%%%%%%%%%%%%%%%%%%%%%%%%%%%%%%%%%%%%%%%%%%%%%%%%%%%
%%%%%%%%%%%%%%%%%%%%%%%%%%%%%%%%%%%%%%%%%%%%%%%%%%%%%%%%%%%%%%%%%%%%%%%%%%%%%%%%%%%%%%%%%%%%%%%

\subsection{Diagonalization\label{section_eom}}
\subsubsection{Electric screening mass\label{section_eom_el}}

Let us consider the Chern--Simons action in Eq. (\ref{actionCS2}).
As it was pointed out in the previous Subsection
(\ref{section_diagrams}), the contribution of the Chern--Simons
action to the equations of motion is reduced to a mixing between
the axial and the vector fields, see Eq. (\ref{action_mixing}).
The relevant part of the whole action involving the $V_0$ and
$A_3$ fields assumes the form: \ba S=\frac{N_c}{12\pi^2R^4}\int
d^4x dr\ \lbr r^3 \lb \dr V_0(r) \rb^2 -  r^3f_{BH}(r)\lb \dr
A_3(r) \rb^2 \right.\nonumber
\\ - \left. 6e_qBR^4 A_3(r) \dr V_0(r) + 2e_qBR^4\dr \lb A_3(r)
V_0(r)\rb \rbr.\label{action_el} \ea The corresponding equations
of motion are:
\ba -\dr(r^3\dr V_0(r))+3e_qBR^4 \dr A_3(r) &=& 0;\label{eqn_vector_mixed}\\
 \dr(r^3f_{BH}(r)\dr A_3(r))-3e_qBR^4 \dr V_0(r) &=& 0.\label{eqn_axial_mixed}\ea
As for the boundary conditions (see Section \ref{section_b=0}),
the values of the gauge fields at the AdS boundary ($r=\infty$) are
determined by the sources (\ref{bc_infty}), where we put $j=0$ and
keep only a source for $V_0$: $V_0(\infty)=\mu,\ A_3(\infty)=0$.
Boundary conditions at the black hole horizon are determined by
Eqs. (\ref{bc_regularity}, \ref{bc_zero}).

A general solution to Eqs. (\ref{eqn_vector_mixed},
\ref{eqn_axial_mixed}) is: \ba V_0(r) &=& \int\limits^r_{\infty}
dr'\, (C_1 P_{\nu}(r_0^2/r'^{2}) + C_2
Q_{\nu}(r_0^2/r'^{2})) + C_3,\label{vsolution}\\
A_3(r) &=& \frac{C_1}{\beta}\ P_{\nu}(r_0^2/r^{2}) +
\frac{C_2}{\beta}\ Q_{\nu}(r_0^2/r^{2}) + C_4\label{asolution},
\ea where $\beta=3e_qBR^4$,
$\nu=-\dfrac{1-\sqrt{1-\beta^2/r_0^4}}{2}$, $P_{\nu}(z)$ and
$Q_{\nu}(z)$ are the Legendre functions of the first and second
order respectively that are single-valued and regular for $|z|<1$.
In our case $\nu$ is real, varies from $0$ ($B=0$) to $-1/2$
($e_qB=\pi^2T^2/6$) and acquires an imaginary part for greater
values of the magnetic field. Let us note that in the case of
$x\in \mathbb{R}$ and $x\rightarrow 1$ $P_{\nu}(x)$ has a finite
limit, while $Q_{\nu}(x)$ possesses a logarithmic singularity
(which corresponds to a branching point in the complex plane).

As one can see, the argument $x$ of the Legendre functions in Eq.
(\ref{vsolution}) varies from $0$ to $1$, the former corresponding
to the AdS boundary and the latter -- to the BH horizon. Therefore
in order to have an axial field, regular at the horizon and zero
at the boundary, according to Eqs. (\ref{bc_regularity},
\ref{bc_zero}, \ref{vsolution}, \ref{asolution}) we should leave
only the Legendre function of the first order in the solutions
(\ref{vsolution}, \ref{asolution}): \be C_2=0. \label{setQ}\ee

Values of the coefficients $C_3, C_4$ in (\ref{vsolution},
\ref{asolution}) are determined by the boundary conditions at the
AdS boundary: \ba C_3=\mu,\ C_4 = \frac{C_1}{\beta}\
P_{\nu}(0),\label{setMu}\ea while the coefficient $C_1$ is to be
determined by the boundary conditions at the horizon
(\ref{bc_regularity}, \ref{bc_zero}): \ba 0=V_0(r_0)
%\lb\rho=1/2\sqrt{\mathcal{M}}\rb=
%\mu-\int\limits_0^{1/2\sqrt{\mathcal{M}}} d\rho\ v(\rho) = \mu -
%\int\limits_0^{1/2\sqrt{\mathcal{M}}} d\rho\ C_1
%P_{\nu}(2\sqrt{\mathcal{M}}\rho)\\ &=& \mu -
%\frac{C_1}{2\sqrt{\mathcal{M}}}\int\limits_0^1 dx\ P_{\nu}(x)
=\mu
- \frac{C_1}{2 r_0^2} \ P_{\nu}^{-1}(0) \Rightarrow C_1
= \mu\ \frac{ 2 r_0^2 } { P_{\nu}^{-1}(0) }.
\label{setP}\ea

Thus, combining Eqs. (\ref{vsolution}, \ref{asolution},
\ref{setQ}, \ref{setMu}, \ref{setP}), we find that the gauge
fields have the following dependence on the radial coordinate $r$:
\ba V_0(r)&=&
%\mu\lb 1 - \frac{2\sqrt{\mathcal{M}}}
%{P^{-1}_{\nu}(0)}\ \int\limits_0^{1/2r^2} d\rho
%P_{\nu}(2\sqrt{\mathcal{M}}\rho) \rb =
%\mu \lb 1 -
%\frac{1}{P^{-1}_{\nu}(0)}\int\limits_0^{\sqrt{\mathcal{M}}/r^2}
%dx\ P_{\nu}(x) \rb \nonumber
= \frac{\mu}{\nu\ P^{-1}_{\nu}(0)}\lb
\frac{r_0^2}{r^2}\
P_{\nu}\left(r_0^2/r^2\right)
- P_{\nu+1}\left(r_0^2/r^2\right) \rb, \label{vexpression}\\
A_3(r)&=&
%2\mu\sqrt{\frac{\mathcal{M}}{\beta^2}}\
%\frac{1}{P^{-1}_{\nu}(0)}\lb P_{\nu}(\sqrt{\mathcal{M}}/r^2) -
%P_{\nu}(0) \rb =
\frac{\mu}{P^{-1}_{\nu}(0)\sqrt{-\nu(\nu+1)}}\
\lb P_{\nu}(r_0^2/r^2) - P_{\nu}(0) \rb.
\label{aexpression}\ea

Taking into account the equations of motion -- Eqs.
(\ref{eqn_vector_mixed}, \ref{eqn_axial_mixed}) -- we get from Eq.
(\ref{action_el}): \ba S= \ld\frac{N_c}{12\pi^2R^4}\int d^4x \lbr
r^3V_0(r)\dr V_0(r) - f_{BH}(r)r^3A_3(r)\dr A_3(r)
-\frac{\beta}{3}\
A_3(r)V_0(r)\rbr\rv^{r=\infty}_{r=r_0}.\label{action_4d} \ea
According to the boundary conditions at the AdS boundary and at
the BH horizon (\ref{bc_regularity}, \ref{bc_zero}) the
second and the third in Eq. (\ref{action_4d}) do not contribute at all, while the first one
is nonzero only at the boundary $r=\infty$. Hence, \ba S&=&
\frac{N_c}{12\pi^2R^4} \lim\limits_{r\rightarrow\infty} \lb
r^3V_0(r)\dr V_0(r)\rb = \frac{N_c}{12\pi^2R^4}\ 2r_0^2\mu^2 \
\frac{P_{\nu}(0)}{P^{-1}_{\nu}(0)} \nonumber
%\\
%&=&\mu^2\frac{N_c}{12}\ T^2\ \frac{2\Gamma\lb 1-\nu/2 \rb\Gamma\lb
%3/2+\nu/2 \rb}{\Gamma\lb 1+\nu/2 \rb\Gamma\lb 1/2-\nu/2\rb}
.\ea Denoting the factor \be F(\nu) \equiv
\frac{P_{\nu}(0)}{P^{-1}_{\nu}(0)} = \frac{2\Gamma\lb 1-\nu/2
\rb\Gamma\lb 3/2+\nu/2 \rb}{\Gamma\lb 1+\nu/2 \rb\Gamma\lb
1/2-\nu/2\rb} \label{F_nu} \ee we obtain the following exact
analytical expression for the Debye mass in any external magnetic
field:

\be m_D^2 = e_q^2\frac{N_c}{3}\ T^2\ F\lb-\frac{1}{2} +
\frac{1}{2}
\sqrt{1-\frac{9e_q^2B^2}{\pi^4T^4}}\rb.\label{el_mass_result}\ee

It is instructive to calculate this quantity numerically in the
case of a quark-gluon plasma that is created during heavy-ion
collisions at RHIC and at the LHC. If we use numerical values
$T\approx 2T_c = 330\pm 20\ {\rm MeV}$ \cite{Kolb} and
$|eB|\approx m^2_{\pi}\approx 2\times 10^4\  {\rm MeV}^2$
\cite{Kharzeev} for RHIC and $T\approx 4-5\cdot T_c = 750\pm 120\
{\rm MeV}$ \cite{Andersen} and $|eB|\approx 15 m^2_{\pi} \approx
3\times 10^5\ {\rm MeV}^2$ \cite{Skokov} for the LHC, we obtain \
\ba \label{RHIC-LHC}
m^2_D &=& (82 \pm 3)^2 \mbox{ MeV$^2$ at RHIC and}\\
\notag m^2_D &=& (185 \pm 35)^2 \mbox{ MeV$^2$ at the LHC.} \ea

In the case of a {\bf weak magnetic field} $e_qB\ll T^2$ it is useful to
expand the function $F(\nu)$ (\ref{F_nu}) in series of $\nu$ and
%\ba F(z) &=& 1 +(1- 2\ln(2)) z +\lb2\ln^2(2)-2\ln(2)\rb z^2
%+\lb 2\ln^2(2)-\frac{4}{3}\ln^3(2)-\frac{1}{2}\ \zeta(3)\rb z^3\nonumber\\
%&+&\lb  \zeta(3) \ln(2)-\frac{1}{2}\
%\zeta(3)+\frac{2}{3}\ln^4(2)-\frac{4}{3}\ \ln^3(2) \rb z^4+O(z^5)
%.\ea
take into account that $\nu=-\frac{1}{2} + \frac{1}{2}
\sqrt{1-\frac{9e_q^2B^2}{\pi^4T^4}}$:

\ba F\lb-\frac{1}{2} + \frac{1}{2}
\sqrt{1-\frac{9e_q^2B^2}{\pi^4T^4}}\rb
%= &1&
%+ \frac{9\left(2\ln(2)-1\right)}{\pi^4}\frac{e_q^2B^2}{T^4} +
%\frac{81\left(2\ln^2(2)-1\right)}{2\pi^8}\frac{e_q^4B^4}{T^8}
%+O\left( \frac{e_q^6B^6}{T^{12}} \right)\nonumber\\
\approx &1& + 0.008923\frac{e_q^2B^2}{T^4} - 0.000021
\frac{e_q^4B^4}{T^8} + O\left(\frac{e_q^6B^6}{T^{12}}\right).
\label{small_B} \ea

%\ba && F\lb-\frac{1-\sqrt{1-\frac{36e_q^2B^2}{\pi^4T^4}}}{2}\rb =
%1 + \frac{9\left(2\ln(2)-1\right)}{\pi^4}\frac{e_q^2B^2}{T^4} +
%\frac{81\left(2\ln^2(2)-1\right)}{2\pi^8}\frac{e_q^4B^4}{T^8}\nonumber\\
%&+& \frac{243\left( 3\zeta(3)+8\ln^3(2)+12\ln^2(2)-12
%\right)}{2\pi^{12}}\frac{e_q^6B^6}{T^{12}}\nonumber\\
%&+&\frac{2187 \left(3\zeta(3)\ln(2)+3\zeta(3)+ 12\ln^2(2)+
%8\ln^3(2)+2\ln^4(2)-15
%\right)}{\pi^{16}} \frac{e_q^8B^8}{T^{16}} + O\lb \frac{e_q^{10}B^{10}}{T^{20}}\rb \nonumber\\
%&\approx& 1 + 0.035691219 \frac{e_q^2B^2}{T^4} - 0.000333730
%\frac{e_q^4B^4}{T^8} + 0.000004707 \frac{e_q^6B^6}{T^{12}} -
%0.000000071 \frac{e_q^8B^8}{T^{16}} + O\lb
%\frac{e_q^{10}B^{10}}{T^{20}}\rb .\ea

%%%%%%%%%%%%%%%%%%%%%%%%%%%%%%%%%%%%%%%%%%%%%%%%%%%%%%%%%%%%%%%%%%%%%%%%%%%%%%%%%%%%%%%%%%%%%%%
%%%%%%%%%%%%%%%%%%%%%%%%%%%%%%%%%%%%%%%%%%%%%%%%%%%%%%%%%%%%%%%%%%%%%%%%%%%%%%%%%%%%%%%%%%%%%%%
%%%%%%%%%%%%%%%%%%%%%%%%%%%%%%%%%%%%%%%%%%%%%%%%%%%%%%%%%%%%%%%%%%%%%%%%%%%%%%%%%%%%%%%%%%%%%%%
%%%%%%%%%%%%%%%%%%%%%%%%%%%%%%%%%%%%%%%%%%%%%%%%%%%%%%%%%%%%%%%%%%%%%%%%%%%%%%%%%%%%%%%%%%%%%%%
%%%%%%%%%%%%%%%%%%%%%%%%%%%%%%%%%%%%%%%%%%%%%%%%%%%%%%%%%%%%%%%%%%%%%%%%%%%%%%%%%%%%%%%%%%%%%%%
%%%%%%%%%%%%%%%%%%%%%%%%%%%%%%%%%%%%%%%%%%%%%%%%%%%%%%%%%%%%%%%%%%%%%%%%%%%%%%%%%%%%%%%%%%%%%%%
%%%%%%%%%%%%%%%%%%%%%%%%%%%%%%%%%%%%%%%%%%%%%%%%%%%%%%%%%%%%%%%%%%%%%%%%%%%%%%%%%%%%%%%%%%%%%%%

\begin{figure}[h]
\centerline{\includegraphics[width=0.4\linewidth]{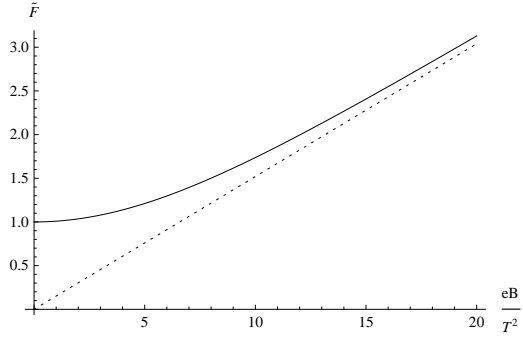}}
\caption{The function $\tilde{F}=F\lb-\frac{1}{2} + \frac{1}{2}
\sqrt{1-\frac{9e_q^2B^2}{\pi^4T^4}}\rb$ (solid) vs its strong
field asymptotics (dashed).} \label{fig_plot}
\end{figure}

In the case of a \textbf{strong magnetic field} we can use the
asymptotic behavior of the gamma function $\Gamma(z) \sim
\sqrt{2\pi}e^{-z}z^{z-1/2}, |z|\rightarrow\infty$, and get \ba
F\lb-\frac{1}{2} + \frac{1}{2}
\sqrt{1-\frac{9e_q^2B^2}{\pi^4T^4}}\rb &\sim&
\frac{3}{2\pi^2}\frac{|e_qB|}{T^2}\lb 1+\frac{17\pi^4}{216}
\frac{T^4}{e_q^2B^2}+O\lb \frac{T^8}{e_q^4B^4} \rb \rb. \ea Thus
in the limit $e_qB\gg T^2$ the Debye mass turns out to be linear
in the magnetic field, in a nice agreement with a weak coupling
result in QED (see \cite{Alexandre}): \be m_D^2 =
e_q^2\frac{N_c}{2\pi^2}\ |e_qB|.\label{el_mass_strongb} \ee The
dependence of the mass on the magnetic field is plotted on Fig.
\ref{fig_plot}. The similarity of the dynamics of strongly coupled
QCD and weakly coupled QED in large external magnetic fields is a
nontrivial phenomenon, which was observed also in
\cite{Son-Thompson}.

%%%%%%%%%%%%%%%%%%%%%%%%%%%%%%%%%%%%%%%%%%%%%%%%%%%%%%%%%%%%%%%%%%%%%%%%%%%%%%%%%%%%%%%%%%%%%%%
%%%%%%%%%%%%%%%%%%%%%%%%%%%%%%%%%%%%%%%%%%%%%%%%%%%%%%%%%%%%%%%%%%%%%%%%%%%%%%%%%%%%%%%%%%%%%%%
%%%%%%%%%%%%%%%%%%%%%%%%%%%%%%%%%%%%%%%%%%%%%%%%%%%%%%%%%%%%%%%%%%%%%%%%%%%%%%%%%%%%%%%%%%%%%%%
%%%%%%%%%%%%%%%%%%%%%%%%%%%%%%%%%%%%%%%%%%%%%%%%%%%%%%%%%%%%%%%%%%%%%%%%%%%%%%%%%%%%%%%%%%%%%%%
%%%%%%%%%%%%%%%%%%%%%%%%%%%%%%%%%%%%%%%%%%%%%%%%%%%%%%%%%%%%%%%%%%%%%%%%%%%%%%%%%%%%%%%%%%%%%%%
%%%%%%%%%%%%%%%%%%%%%%%%%%%%%%%%%%%%%%%%%%%%%%%%%%%%%%%%%%%%%%%%%%%%%%%%%%%%%%%%%%%%%%%%%%%%%%%
%%%%%%%%%%%%%%%%%%%%%%%%%%%%%%%%%%%%%%%%%%%%%%%%%%%%%%%%%%%%%%%%%%%%%%%%%%%%%%%%%%%%%%%%%%%%%%%

\subsubsection{Magnetic screening mass \label{section_eom_mag}}

The calculation of the magnetic screening mass (\ref{mag_mass}) is
quite similar to one in the previous subsection. The only dynamical fields we need to
consider are $V_3(0)$ and $A_0(r)$, the former having a source $j$
with respect to which we have to variate the action twice in order
to get the two-point correlator $\langle J_3 J_3 \rangle$, and the
mixing due to the presence of the CS action
proportional to the external magnetic field.

The action we are dealing with in this subsection is the following
(see Eq. (\ref{action_mixing})): \ba S=\frac{N_c}{12\pi^2R^4}\int
d^4x dr\ \lbr r^3 \lb \dr A_0(r) \rb^2 -  r^3f_{BH}(r)\lb \dr
V_3(r) \rb^2 \right.\nonumber
\\ + \left. 6e_qBR^4 V_3(r) \dr A_0(r) - 2e_qBR^4\dr \lb V_3(r)
A_0(r)\rb \rbr.\label{action_mag} \ea It generates the equations
of motion analogous to Eqs. (\ref{eqn_axial_mixed},
\ref{eqn_vector_mixed}):
\ba -\dr(r^3\dr A_0(r))+3e_qBR^4 \dr V_3(r) &=& 0;\label{eqn_vector_mag}\\
 \dr(r^3f_{BH}(r)\dr V_3(r))-3e_qBR^4 \dr A_0(r) &=& 0.\label{eqn_axial_mag}\ea
The boundary conditions in this case are determined by the fact
that $V_3$ has a source $j$ at the AdS boundary while $A_0$ has
none, and at the horizon --- by Eqs. (\ref{bc_regularity},
\ref{bc_zero}). The only solution to Eqs. (\ref{eqn_axial_mag},
\ref{eqn_vector_mag}) with these boundary conditions is: \be
A_0(r)\equiv 0,\ V_3(r)\equiv j.\label{solution_mag} \ee The
corresponding action (\ref{action_mag}) is zero, therefore the
magnetic screening mass is zero: \be m_{D\
Mag}=0\label{mag_mass_ads}\ee even in the presence of the magnetic
field, while as already mentioned in \cite{Blaizot} it is shown to
vanish to all orders of perturbation theory in the absence of an
external field. Our result is obtained in the leading order of
$\frac{1}{N_c}$ expansion and, as in (\ref{O(1)}), can acquire
subleading corrections. Nevertheless, let us note, that in QED
($N_c=1$) a similar feature takes place:  $m_{D \ Mag}=0$ in an
external magnetic field \cite{Alexandre}.

%%%%%%%%%%%%%%%%%%%%%%%%%%%%%%%%%%%%%%%%%%%%%%%%%%%%%%%%%%%%%%%%%%%%%%%%%%%%%%%%%%%%%%%%%%%%%%%
%%%%%%%%%%%%%%%%%%%%%%%%%%%%%%%%%%%%%%%%%%%%%%%%%%%%%%%%%%%%%%%%%%%%%%%%%%%%%%%%%%%%%%%%%%%%%%%
%%%%%%%%%%%%%%%%%%%%%%%%%%%%%%%%%%%%%%%%%%%%%%%%%%%%%%%%%%%%%%%%%%%%%%%%%%%%%%%%%%%%%%%%%%%%%%%
%%%%%%%%%%%%%%%%%%%%%%%%%%%%%%%%%%%%%%%%%%%%%%%%%%%%%%%%%%%%%%%%%%%%%%%%%%%%%%%%%%%%%%%%%%%%%%%
%%%%%%%%%%%%%%%%%%%%%%%%%%%%%%%%%%%%%%%%%%%%%%%%%%%%%%%%%%%%%%%%%%%%%%%%%%%%%%%%%%%%%%%%%%%%%%%
%%%%%%%%%%%%%%%%%%%%%%%%%%%%%%%%%%%%%%%%%%%%%%%%%%%%%%%%%%%%%%%%%%%%%%%%%%%%%%%%%%%%%%%%%%%%%%%
%%%%%%%%%%%%%%%%%%%%%%%%%%%%%%%%%%%%%%%%%%%%%%%%%%%%%%%%%%%%%%%%%%%%%%%%%%%%%%%%%%%%%%%%%%%%%%%

\subsection{Lower temperature case in holography}

As we move to lower temperatures, expression in Eq.
(\ref{el_mass_result}) asymptotically tends to Eq.
(\ref{el_mass_strongb}) and thus the Debye mass grows linearly
with the magnetic field. However, this result holds only if we
impose the same boundary conditions in the infrared region of the
AdS as in the case of high temperatures. In reality, the geometry
of this region may change drastically when temperature $T$
approaches the $\Lambda_{QCD}$ scale, undergoing the Hawking--Page
transition associated with the deconfinement phase transition of
QCD \cite{Witten}. One of AdS/QCD models
\cite{Erlich-Katz-Son-Stephanov} suggests, for instance, that in
the confinement phase one places a hard wall at a certain point
$r=r_m$ and impose a Neumann boundary condition for all fields at
$r=r_m$, where $r_m \propto \Lambda_{QCD}^{-1}$: \be \dr
V_{\mu}(r_m)=\dr A_{\nu}(r_m)=0,\label{bc_neumann}\ee where
$\mu,\nu=0,1,2,3$. This hard wall is situated inside the BH at
high temperatures and is uncovered by the BH horizon when the
temperature drops lower than $\approx T_c\sim\Lambda_{QCD}$. If we
consider the equations of motion (\ref{eqn_axial_mixed},
\ref{eqn_vector_mixed}) with the new Neumann boundary conditions
(\ref{bc_neumann}), a simple analysis demonstrates that their only
possible solution is a pair of identically zero functions:
$A_3(r)=V_0(r)\equiv 0$. Thus the Debye mass at temperatures below
the phase transition appears to be zero in this particular model.
The same holds true for the magnetic screening mass.

However, this result strongly depends on the type of the boundary
conditions that we impose at the infrared boundary and may only be
considered as a qualitative indication. We shall
treat the case of a zero temperature more rigorously in the next
section.

%%%%%%%%%%%%%%%%%%%%%%%%%%%%%%%%%%%%%%%%%%%%%%%%%%%%%%%%%%%%%%%%%%%%%%%%%%%%%%%%%%%%%%%%%%%%%%%
%%%%%%%%%%%%%%%%%%%%%%%%%%%%%%%%%%%%%%%%%%%%%%%%%%%%%%%%%%%%%%%%%%%%%%%%%%%%%%%%%%%%%%%%%%%%%%%
%%%%%%%%%%%%%%%%%%%%%%%%%%%%%%%%%%%%%%%%%%%%%%%%%%%%%%%%%%%%%%%%%%%%%%%%%%%%%%%%%%%%%%%%%%%%%%%
%%%%%%%%%%%%%%%%%%%%%%%%%%%%%%%%%%%%%%%%%%%%%%%%%%%%%%%%%%%%%%%%%%%%%%%%%%%%%%%%%%%%%%%%%%%%%%%
%%%%%%%%%%%%%%%%%%%%%%%%%%%%%%%%%%%%%%%%%%%%%%%%%%%%%%%%%%%%%%%%%%%%%%%%%%%%%%%%%%%%%%%%%%%%%%%
%%%%%%%%%%%%%%%%%%%%%%%%%%%%%%%%%%%%%%%%%%%%%%%%%%%%%%%%%%%%%%%%%%%%%%%%%%%%%%%%%%%%%%%%%%%%%%%
%%%%%%%%%%%%%%%%%%%%%%%%%%%%%%%%%%%%%%%%%%%%%%%%%%%%%%%%%%%%%%%%%%%%%%%%%%%%%%%%%%%%%%%%%%%%%%%

\section{Confinement phase\label{section_chpt}}

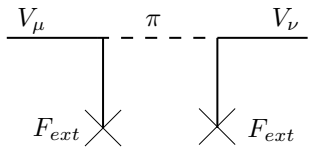
\begin{figure*}[h!]
%\center{\includegraphics{figure3.eps}}
\input{diagram3.pstex_t}
\caption{The correction to
the photon polarization operator in the external field in
ChPT}\label{fig_chpt}
\end{figure*}

To study the screening masses in the confinement phase we make use
of the Chiral Perturbation Theory (ChPT) \cite{ChPT}. As was shown
in the previous sections, in the holographic approach the whole
effect is governed by a Chern--Simons type interaction.
Interestingly enough, in the chiral perturbation theory there
exists a quite similar diagram, describing the correlation
function of two vector currents in an external field (Fig.
\ref{fig_chpt}). It includes two anomalous vertices and an
exchange of a $\pi^0$ meson.

Let us emphasize that the anomalous vertices correspond to the
Goldstone--Wilczek currents emerging upon the computation of the
fermionic loop in the varying meson field. That is, evaluating the
Debye screening we actually look at the correlation of  two
induced electric charges if  simultaneously magnetic and $\pi_0$
meson fields are switched on. To some extent this is a kind of a
contribution which is quadratic in the effective chiral chemical
potential.

As we are interested mainly in the
effects of an external magnetic field, we restrict ourselves to
the case of zero temperature. The relevant terms of ChPT Lagrangian
in the external field are \cite{ChPT}: \be L_{\chi PT} =
\frac{1}{2} \p_\mu \phi^{\dag} \p^\mu \phi + \frac{1}{2} M_\pi^2
\phi^{\dag} \phi - \frac{\alpha_{em}}{4 \pi} \frac{1}{f_\pi} \phi
F_{\mu \nu} \tilde{F}_{\mu \nu} \ee As was mentioned above, the
correction to the photon polarization operator in the external
field arises already at the tree level and can be computed quite
easily.

\noindent
It reflects the well-known photon-pion mixing in the magnetic field.

In the case when $B_{ext}=(0,0,B)$ one gets the
following result for $\Pi_{\mu \nu}(q,\omega)$ from the diagram
(Fig. \ref{fig_chpt}):
\begin{align}
\label{ChPT_el}
\Pi_{00}(q,\omega) &= \frac{\alpha_{em}^2}{(4 \pi)^2} \frac{B^2}{f_\pi^2} \frac{-q_3 q_3}{\omega^2 - |\vec{q}|^2 - M_{\pi}^2} \\
\notag \Pi_{33}(q,\omega) &= \frac{\alpha_{em}^2}{(4 \pi)^2}
\frac{B^2}{f_\pi^2} \frac{-\omega^2}{\omega^2 - |\vec{q}|^2 -
M_{\pi}^2}
\end{align}
To obtain the screening masses (\ref{el_mass}),(\ref{mag_mass}),
we need to set $\omega=0$ and then take $\vec{q}^2 = - m_D^2$.
Then the magnetic mass associated with $\Pi_{33}$ vanishes,
coinciding with the result of the holographic calculation. However the
Debye mass behaves much more interestingly. Suppose we consider
the correlator with finite but small momentum $|\vec{q}|^2 = -
m_D^2$. The 3$^{{\rm rd}}$ component of the momentum can be
expressed as $q_3 = m_D \ cos(\widehat{\vec{q},\vec{B}})$. Then
the equation (\ref{ChPT_el}) yields: \be m_D^2 \left(m_D^2 -
M_\pi^2 - \frac{\alpha_{em}^2}{(4\pi)^2} \frac{B^2}{f_\pi^2}\
\cos^2(\widehat{\vec{q},\vec{B}}) \right) = 0. \notag \ee

Naively this equation has two solutions, $m_D=0$ and
$m_D^2=M_\pi^2 + \dfrac{\alpha_{em}^2}{(4\pi)^2}
\dfrac{B^2}{f_\pi^2}\ \cos^2(\widehat{\vec{q},\vec{B}})$. The
second one arises due to the fact that the definition of the Debye
screening in the confining phase in the magnetic field needs for
some care. In the presence of a magnetic field in the confinement
phase the photon mixes with the pion due to the anomaly. Therefore
the Debye mass is naturally defined upon the diagonalization of
two mixing states; one has to look for the poles of the propagator
of the states. One of the poles $m_D^2=0$ corresponds to the Debye
mass of the photon, while another reflects the shift of the pion
mass due to an admixture of the photon, and therefore this pole is
irrelevant here. Hence we get a vanishing anomalous contribution
at zero temperature which seems quite natural.

%The second one seems to be unphysical since the definition of the
%Debye screening in the confining phase in the magnetic field
%contrary to the deconfinement phase needs for some care. As we
%have mentioned the photon in the confinement phase get mixed with
%the pion in the magnetic field therefore the Debye mass is
%naturally defined upon the diagonalization of two mixing states
%only. Upon diagonalization the pion mass contribution does not
%work hence in the real world we get a small or vanishing anomalous
%contribution at zero temperature which seems quite natural.

Nevertheless let us point out that in the case when the pion is
massless Eq. (\ref{ChPT_el}) has only one solution
$m_D^2=\dfrac{\alpha_{em}^2}{(4\pi)^2} \dfrac{B^2}{f_\pi^2}\
\cos^2(\widehat{\vec{q},\vec{B}})$ which indicates the existence
of an anisotropic deformation of the Coulomb potential at zero
temperature in an external magnetic field. However let us stress
that we have discussed in this Section the anomalous contribution
only, while there is an additional contribution to the
polarization operator of a loop with charged pions.

%%%%%%%%%%%%%%%%%%%%%%%%%%%%%%%%%%%%%%%%%%%%%%%%%%%%%%%%%%%%%%%%%%%%%%%%%%%%%%%%%%%%%%%%%%%%%%%
%%%%%%%%%%%%%%%%%%%%%%%%%%%%%%%%%%%%%%%%%%%%%%%%%%%%%%%%%%%%%%%%%%%%%%%%%%%%%%%%%%%%%%%%%%%%%%%
%%%%%%%%%%%%%%%%%%%%%%%%%%%%%%%%%%%%%%%%%%%%%%%%%%%%%%%%%%%%%%%%%%%%%%%%%%%%%%%%%%%%%%%%%%%%%%%
%%%%%%%%%%%%%%%%%%%%%%%%%%%%%%%%%%%%%%%%%%%%%%%%%%%%%%%%%%%%%%%%%%%%%%%%%%%%%%%%%%%%%%%%%%%%%%%
%%%%%%%%%%%%%%%%%%%%%%%%%%%%%%%%%%%%%%%%%%%%%%%%%%%%%%%%%%%%%%%%%%%%%%%%%%%%%%%%%%%%%%%%%%%%%%%
%%%%%%%%%%%%%%%%%%%%%%%%%%%%%%%%%%%%%%%%%%%%%%%%%%%%%%%%%%%%%%%%%%%%%%%%%%%%%%%%%%%%%%%%%%%%%%%
%%%%%%%%%%%%%%%%%%%%%%%%%%%%%%%%%%%%%%%%%%%%%%%%%%%%%%%%%%%%%%%%%%%%%%%%%%%%%%%%%%%%%%%%%%%%%%%

\section{Conclusion\label{section_conclusion}}

In this work we have studied the yet unexplored corrections to
electromagnetic screening masses in a deconfined QCD plasma due to
strong interactions. At temperatures, larger than the temperature
of deconfinement we have used the holographic AdS/QCD model to
describe the QCD dynamics. The advantage of the holographic
results we have obtained (\ref{el_mass_result},
\ref{mag_mass_ads}) is the exact treatment of the external
magnetic field, namely, all orders of perturbation theory have
been summed up. Having an analytical formula for the Debye
screening mass (\ref{el_mass_result}), we have studied various
limits in the external field. Given that the external field is
small we have found that the non-perturbative result
(\ref{el_mass_b0}, \ref{small_B}) equals that of the first order
of the perturbation theory in QED (\ref{one-loop}) \cite{Weldon}.
The behavior of the electric mass in a large magnetic field
(\ref{el_mass_strongb}) coincides with a result of a calculation
in one-loop QED in an external field and at nonzero temperature
\cite{Alexandre}. We have also found the magnetic screening mass
to be zero at any values of the magnetic field.

It turned out that in our holographic model the dependence of the
screening mass on the magnetic field is fully driven by the
Chern--Simons term. Motivated by this fact, we have studied a
similar diagram in the chiral perturbation theory and found an
interesting anisotropy of the Debye mass in the magnetic field.

The obtained results show a nice agreement with all previous
studies of the Debye screening and demonstrate the sensibility of
the holographic model considered in this paper. It would be very
natural to extend our Debye mass consideration to a dense QCD. In
the deconfined phase such setup is holographically described by a
charged black hole. On the other hand in the confining phase there
are arguments that in a large magnetic field matter behaves as a
stack of pionic domain walls \cite{son}. Such an unusual state is
stable both perturbatively \cite{son} and non-perturbatively
\cite{gv}. It would be very interesting to investigate the
screening behavior in this phase as well.

\acknowledgments

We would like to thank V.~I.~Shevchenko for the question initiated
this research and A.~V.~Zayakin for fruitful discussions. Research
of P.~N.~K. was supported by the Dynasty Foundation, the grant
RFBR-09-02-00308 and by the Ministry of Education and Science of
the Russian Federation under contract 14.740.11.0081. Research of
A.~K. was supported by the Dynasty Foundation, the grant
RFBR-10-02-01483 and by the Ministry of Education and Science of
the Russian Federation under contract 14.740.11.0347. The work of
A.~G. is supported in part by the grants PICS- 07-0292165,
RFBR-09-02-00308 and CRDF - RUP2-2961-MO-09. A.~G. thanks IPhT
at Sacley where the part of this work has been done
for the hospitality and support.

%%%%%%%%%%%%%%%%%%%%%%%%%%%%%%%%%%%%%%%%%%%%%%%%%%%%%%%%%%%%%%%%%%%%%%%%%%%%%%%%%%%%%%%%%%%%%%%
%%%%%%%%%%%%%%%%%%%%%%%%%%%%%%%%%%%%%%%%%%%%%%%%%%%%%%%%%%%%%%%%%%%%%%%%%%%%%%%%%%%%%%%%%%%%%%%
%%%%%%%%%%%%%%%%%%%%%%%%%%%%%%%%%%%%%%%%%%%%%%%%%%%%%%%%%%%%%%%%%%%%%%%%%%%%%%%%%%%%%%%%%%%%%%%
%%%%%%%%%%%%%%%%%%%%%%%%%%%%%%%%%%%%%%%%%%%%%%%%%%%%%%%%%%%%%%%%%%%%%%%%%%%%%%%%%%%%%%%%%%%%%%%
%%%%%%%%%%%%%%%%%%%%%%%%%%%%%%%%%%%%%%%%%%%%%%%%%%%%%%%%%%%%%%%%%%%%%%%%%%%%%%%%%%%%%%%%%%%%%%%
%%%%%%%%%%%%%%%%%%%%%%%%%%%%%%%%%%%%%%%%%%%%%%%%%%%%%%%%%%%%%%%%%%%%%%%%%%%%%%%%%%%%%%%%%%%%%%%
%%%%%%%%%%%%%%%%%%%%%%%%%%%%%%%%%%%%%%%%%%%%%%%%%%%%%%%%%%%%%%%%%%%%%%%%%%%%%%%%%%%%%%%%%%%%%%%

\end{document}

%% file: diagram1a.pstex_t
\begin{picture}(0,0)%
\includegraphics{diagram1a.pstex}%
\end{picture}%
\setlength{\unitlength}{3315sp}%
\begingroup\makeatletter\ifx\SetFigFontNFSS\undefined%
\gdef\SetFigFontNFSS#1#2#3#4#5{%
  \reset@font\fontsize{#1}{#2pt}%
  \fontfamily{#3}\fontseries{#4}\fontshape{#5}%
  \selectfont}%
\fi\endgroup%
\begin{picture}(8745,1836)(886,-4189)
\put(2116,-2941){\makebox(0,0)[lb]{\smash{{\SetFigFontNFSS{10}{12.0}{\rmdefault}{\mddefault}{\updefault}{\color[rgb]{0,0,0}$A_3$}%
}}}}
\put(1621,-4111){\makebox(0,0)[lb]{\smash{{\SetFigFontNFSS{10}{12.0}{\rmdefault}{\mddefault}{\updefault}{\color[rgb]{0,0,0}$F_{12}$}%
}}}}
\put(2521,-4111){\makebox(0,0)[lb]{\smash{{\SetFigFontNFSS{10}{12.0}{\rmdefault}{\mddefault}{\updefault}{\color[rgb]{0,0,0}$F_{12}$}%
}}}}
\put(901,-2536){\makebox(0,0)[lb]{\smash{{\SetFigFontNFSS{10}{12.0}{\rmdefault}{\mddefault}{\updefault}{\color[rgb]{0,0,0}$V_0$}%
}}}}
\put(3376,-2536){\makebox(0,0)[lb]{\smash{{\SetFigFontNFSS{10}{12.0}{\rmdefault}{\mddefault}{\updefault}{\color[rgb]{0,0,0}$V_0$}%
}}}}
\put(5296,-2896){\makebox(0,0)[lb]{\smash{{\SetFigFontNFSS{10}{12.0}{\rmdefault}{\mddefault}{\updefault}{\color[rgb]{0,0,0}$A_3$}%
}}}}
\put(6151,-2896){\makebox(0,0)[lb]{\smash{{\SetFigFontNFSS{10}{12.0}{\rmdefault}{\mddefault}{\updefault}{\color[rgb]{0,0,0}$V_0$}%
}}}}
\put(7051,-2896){\makebox(0,0)[lb]{\smash{{\SetFigFontNFSS{10}{12.0}{\rmdefault}{\mddefault}{\updefault}{\color[rgb]{0,0,0}$A_3$}%
}}}}
\put(4846,-4111){\makebox(0,0)[lb]{\smash{{\SetFigFontNFSS{10}{12.0}{\rmdefault}{\mddefault}{\updefault}{\color[rgb]{0,0,0}$F_{12}$}%
}}}}
\put(6646,-4066){\makebox(0,0)[lb]{\smash{{\SetFigFontNFSS{10}{12.0}{\rmdefault}{\mddefault}{\updefault}{\color[rgb]{0,0,0}$F_{12}$}%
}}}}
\put(7546,-4066){\makebox(0,0)[lb]{\smash{{\SetFigFontNFSS{10}{12.0}{\rmdefault}{\mddefault}{\updefault}{\color[rgb]{0,0,0}$F_{12}$}%
}}}}
\put(5746,-4111){\makebox(0,0)[lb]{\smash{{\SetFigFontNFSS{10}{12.0}{\rmdefault}{\mddefault}{\updefault}{\color[rgb]{0,0,0}$F_{12}$}%
}}}}
\put(4051,-2536){\makebox(0,0)[lb]{\smash{{\SetFigFontNFSS{10}{12.0}{\rmdefault}{\mddefault}{\updefault}{\color[rgb]{0,0,0}$V_0$}%
}}}}
\put(8326,-2536){\makebox(0,0)[lb]{\smash{{\SetFigFontNFSS{10}{12.0}{\rmdefault}{\mddefault}{\updefault}{\color[rgb]{0,0,0}$V_0$}%
}}}}
\put(9001,-3436){\makebox(0,0)[lb]{\smash{{\SetFigFontNFSS{20}{24.0}{\rmdefault}{\mddefault}{\updefault}{\color[rgb]{0,0,0}. . .}%
}}}}
\end{picture}%

%% file: diagram1b.pstex_t
\begin{picture}(0,0)%
\includegraphics{diagram1b.pstex}%
\end{picture}%
\setlength{\unitlength}{3315sp}%
\begingroup\makeatletter\ifx\SetFigFontNFSS\undefined%
\gdef\SetFigFontNFSS#1#2#3#4#5{%
  \reset@font\fontsize{#1}{#2pt}%
  \fontfamily{#3}\fontseries{#4}\fontshape{#5}%
  \selectfont}%
\fi\endgroup%
\begin{picture}(8745,1836)(886,-4189)
\put(9001,-3436){\makebox(0,0)[lb]{\smash{{\SetFigFontNFSS{20}{24.0}{\rmdefault}{\mddefault}{\updefault}{\color[rgb]{0,0,0}. . .}%
}}}}
\put(3376,-2536){\makebox(0,0)[lb]{\smash{{\SetFigFontNFSS{10}{12.0}{\rmdefault}{\mddefault}{\updefault}{\color[rgb]{0,0,0}$V_3$}%
}}}}
\put(4051,-2536){\makebox(0,0)[lb]{\smash{{\SetFigFontNFSS{10}{12.0}{\rmdefault}{\mddefault}{\updefault}{\color[rgb]{0,0,0}$V_3$}%
}}}}
\put(8326,-2536){\makebox(0,0)[lb]{\smash{{\SetFigFontNFSS{10}{12.0}{\rmdefault}{\mddefault}{\updefault}{\color[rgb]{0,0,0}$V_3$}%
}}}}
\put(5296,-2896){\makebox(0,0)[lb]{\smash{{\SetFigFontNFSS{10}{12.0}{\rmdefault}{\mddefault}{\updefault}{\color[rgb]{0,0,0}$A_0$}%
}}}}
\put(6151,-2896){\makebox(0,0)[lb]{\smash{{\SetFigFontNFSS{10}{12.0}{\rmdefault}{\mddefault}{\updefault}{\color[rgb]{0,0,0}$V_3$}%
}}}}
\put(7051,-2896){\makebox(0,0)[lb]{\smash{{\SetFigFontNFSS{10}{12.0}{\rmdefault}{\mddefault}{\updefault}{\color[rgb]{0,0,0}$A_0$}%
}}}}
\put(4846,-4111){\makebox(0,0)[lb]{\smash{{\SetFigFontNFSS{10}{12.0}{\rmdefault}{\mddefault}{\updefault}{\color[rgb]{0,0,0}$F_{12}$}%
}}}}
\put(6646,-4066){\makebox(0,0)[lb]{\smash{{\SetFigFontNFSS{10}{12.0}{\rmdefault}{\mddefault}{\updefault}{\color[rgb]{0,0,0}$F_{12}$}%
}}}}
\put(7546,-4066){\makebox(0,0)[lb]{\smash{{\SetFigFontNFSS{10}{12.0}{\rmdefault}{\mddefault}{\updefault}{\color[rgb]{0,0,0}$F_{12}$}%
}}}}
\put(5746,-4111){\makebox(0,0)[lb]{\smash{{\SetFigFontNFSS{10}{12.0}{\rmdefault}{\mddefault}{\updefault}{\color[rgb]{0,0,0}$F_{12}$}%
}}}}
\put(2116,-2941){\makebox(0,0)[lb]{\smash{{\SetFigFontNFSS{10}{12.0}{\rmdefault}{\mddefault}{\updefault}{\color[rgb]{0,0,0}$A_0$}%
}}}}
\put(2521,-4111){\makebox(0,0)[lb]{\smash{{\SetFigFontNFSS{10}{12.0}{\rmdefault}{\mddefault}{\updefault}{\color[rgb]{0,0,0}$F_{12}$}%
}}}}
\put(901,-2536){\makebox(0,0)[lb]{\smash{{\SetFigFontNFSS{10}{12.0}{\rmdefault}{\mddefault}{\updefault}{\color[rgb]{0,0,0}$V_3$}%
}}}}
\put(1621,-4111){\makebox(0,0)[lb]{\smash{{\SetFigFontNFSS{10}{12.0}{\rmdefault}{\mddefault}{\updefault}{\color[rgb]{0,0,0}$F_{12}$}%
}}}}
\end{picture}%

%% file: diagram3.pstex_t
\begin{picture}(0,0)%
\includegraphics{diagram3.pstex}%
\end{picture}%
\setlength{\unitlength}{3315sp}%
\begingroup\makeatletter\ifx\SetFigFontNFSS\undefined%
\gdef\SetFigFontNFSS#1#2#3#4#5{%
  \reset@font\fontsize{#1}{#2pt}%
  \fontfamily{#3}\fontseries{#4}\fontshape{#5}%
  \selectfont}%
\fi\endgroup%
\begin{picture}(2294,1095)(2679,-2908)
\put(4681,-1996){\makebox(0,0)[lb]{\smash{{\SetFigFontNFSS{10}{12.0}{\rmdefault}{\mddefault}{\updefault}{\color[rgb]{0,0,0}$V_\nu$}%
}}}}
\put(3736,-1996){\makebox(0,0)[lb]{\smash{{\SetFigFontNFSS{10}{12.0}{\rmdefault}{\mddefault}{\updefault}{\color[rgb]{0,0,0}$\pi$}%
}}}}
\put(2776,-1996){\makebox(0,0)[lb]{\smash{{\SetFigFontNFSS{10}{12.0}{\rmdefault}{\mddefault}{\updefault}{\color[rgb]{0,0,0}$V_\mu$}%
}}}}
\put(2896,-2821){\makebox(0,0)[lb]{\smash{{\SetFigFontNFSS{10}{12.0}{\rmdefault}{\mddefault}{\updefault}{\color[rgb]{0,0,0}$F_{ext}$}%
}}}}
\put(4501,-2830){\makebox(0,0)[lb]{\smash{{\SetFigFontNFSS{10}{12.0}{\rmdefault}{\mddefault}{\updefault}{\color[rgb]{0,0,0}$F_{ext}$}%
}}}}
\end{picture}%

%% file: debye_mass-14-arx.bbl
\begin{thebibliography}{99}

%\cite{Kapusta:2006pm}
\bibitem{Kapusta}
  J.~I.~Kapusta and C.~Gale,
  ``{\it Finite-temperature field theory: Principles and applications},''
%\href{http://www.slac.stanford.edu/spires/find/hep/www?irn=7209002}{SPIRES entry}
Cambridge, UK: Univ. Pr. (2006) 428 p.


\bibitem{Le Bellac}
M.~Le Bellac, ``{\it Thermal Field Theory}'', Cambridge Monographs
on Mathematical Physics (1996) 256 p.

%\cite{Weldon:1982aq}
\bibitem{Weldon}
  H.~A.~Weldon,
  ``Covariant Calculations At Finite Temperature: The Relativistic Plasma,''
  {\it Phys.\ Rev.}\ {\bf D 26}, 1394 (1982).
  %%CITATION = PHRVA,D26,1394;%%

%\cite{Blaizot:1995kg}
\bibitem{Blaizot}
  J.~P.~Blaizot, E.~Iancu and R.~R.~Parwani,
  ``On The Screening Of Static Electromagnetic Fields In Hot QED Plasmas,''
  {\it Phys.\ Rev.}\ {\bf D 52}, 2543 (1995)
  \texttt{ePrint arXiv: hep-ph/9504408}.
  %%CITATION = PHRVA,D52,2543;%%

\bibitem{Kharzeev}
Dmitri E. Kharzeev, Larry D. McLerran, Harmen J. Warringa, ``The
Effects of topological charge change in heavy ion collisions:
'Event by event P and CP violation' ''. {\it Nucl. Phys.} {\bf
A803}, 227-253, 2008; \texttt{e-Print arXiv: 0711.0950 [hep-ph]}.


\bibitem{Skokov}
V. Skokov, A.Yu. Illarionov, V. Toneev, ``Estimate of the magnetic
field strength in heavy-ion collisions'', {\it Int. J. Mod. Phys.}
{\bf A24}, 5925-5932, 2009; \texttt{e-Print arXiv: 0907.1396
[nucl-th]}.



%\cite{Alexandre:2000jc}
\bibitem{Alexandre}
  J.~Alexandre,
  ``Vacuum polarization in thermal QED with an external magnetic field,''
  {\it Phys.\ Rev.}\ {\bf D 63}, 073010 (2001)
  \texttt{e-Print arXiv: hep-th/0009204}.
  %%CITATION = PHRVA,D63,073010;%%

%\cite{Schwinger:1951nm}
\bibitem{Schwinger}
  J.~S.~Schwinger,
  ``On gauge invariance and vacuum polarization,''
  Phys.\ Rev.\  {\bf 82}, 664 (1951).
  %%CITATION = PHRVA,82,664;%%


\bibitem{Maldacena}
J. M. Maldacena, {\it Adv. Theor. Math. Phys.} {\bf 2}, 231-252
(1998), \texttt{e-Print arXiv: hep-th/9711200},\\
S. S. Gubser, I. R. Klebanov and A. M. Polyakov, {\it Phys. Lett.
B} {\bf 428}, 105 (1998), \texttt{e-Print arXiv: hep-th/9802109},\\
E. Witten, {\it Adv. Theor. Math. Phys.} {\bf 2}, 253 (1998),
\texttt{e-Print arXiv: hep-th/9802150}.





%\cite{Witten:1998zw}
\bibitem{Witten}
  E.~Witten,
  ``Anti-de Sitter space, thermal phase transition, and confinement in  gauge
  theories,''
  {\it Adv.\ Theor.\ Math.\ Phys.}\  {\bf 2}, 505 (1998)
  \texttt{e-Print arXiv: hep-th/9803131}.
  %%CITATION = 00203,2,505;%%

%\cite{Son:2002sd}
\bibitem{Son-Starinets}
  D.~T.~Son and A.~O.~Starinets,
  ``Minkowski-space correlators in AdS/CFT correspondence: Recipe and
  applications,''
  {\it JHEP}\ {\bf 0209}, 042 (2002)
  \texttt{e-Print arXiv: hep-th/0205051}.
  %%CITATION = JHEPA,0209,042;%%

%\cite{Hartnoll:2008vx}
\bibitem{Herzog}
  S.~A.~Hartnoll, C.~P.~Herzog and G.~T.~Horowitz,
  ``Building a Holographic Superconductor,''
  {\it Phys.\ Rev.\ Lett.}\  {\bf 101}, 031601 (2008)
  \texttt{e-Print arXiv: 0803.3295 [hep-th]}.
  %%CITATION = PRLTA,101,031601;%%


%\cite{Gasser:1983yg}
\bibitem{ChPT}
  J.~Gasser and H.~Leutwyler,
  ``Chiral Perturbation Theory To One Loop,''
  {\it Annals Phys.}\  {\bf 158}, 142 (1984).
  %%CITATION = APNYA,158,142;%%

%\cite{Bell:1969ts}

\bibitem{Erlich-Katz-Son-Stephanov}
Joshua Erlich, Emanuel Katz, Dam T. Son, Mikhail A. Stephanov,
``QCD and a holographic model of hadrons'', SLAC-PUB-10965,
WM-05-101, INT-PUB-05-02, {\it Phys. Rev. Lett.} {\bf 95}, 261602,
2005; \texttt{e-Print arXiv: hep-ph/0501128}.

\bibitem{SUMrules}
M. A. Shifman, A. I. Vainshtein, V. I. Zakharov, ``QCD and
resonance physics. Theoretical foundations'', {\it Nucl. Phys. B}
{\bf 147}, 385-447 (1979).


%\cite{Witten:1983tw}
\bibitem{Witten-Wess}
  E.~Witten,
  ``Global Aspects Of Current Algebra,''
  {\it Nucl.\ Phys.}\ {\bf B 223}, 422 (1983).
  %%CITATION = NUPHA,B223,422;%%



\bibitem{Yee}
  H.~U.~Yee,
  ``Holographic Chiral Magnetic Conductivity,''
  {\it JHEP} {\bf 0911}, 085 (2009)
  \texttt{e-Print arXiv: 0908.4189 [hep-th]}.
  %%CITATION = JHEPA,0911,085;%%


\bibitem{Chern-Simons}
H.~R.~Grigoryan, A.~V.~Radyushkin, ``Anomalous Form Factor of the
Neutral Pion in Extended AdS/QCD Model with Chern-Simons Term'',
JLAB-THY-08-802, {\it Phys. Rev.} {\bf D77}, 115024, 2008; \texttt{e-Print: arXiv:0803.1143 [hep-ph]}\\
%
A.~S.~Gorsky, A.~A.~Krikun, ``Magnetic susceptibility of the quark
condensate via holography'', ITEP-TH-04-09, {\it Phys. Rev.} {\bf
D79}, 086015, 2009; e-Print: arXiv:0902.1832 [hep-ph];\\
%
A. Rebhan, A. Schmitt and S. A. Stricker, ``Anomalies and the
chiral magnetic effect in the Sakai-Sugimoto model", \textit{JHEP}
\textbf{1001}, 026 (2010); \texttt{arXiv: 0909.4782[hep-th]};\\
%
A.~S.~Gorsky, P.~N.~Kopnin, A.~V.~Zayakin, On the Chiral Magnetic
Effect in Soft-Wall AdS/QCD, \texttt{arXiv: 1003.2293[hep-ph]}.


%\cite{Krikun:2010ip}
\bibitem{Krikun-4point}
  A.~Krikun,
  ``Four-point correlator of vector currents and electric current
  susceptibility in holographic QCD,''
  {\it Phys.\ Lett.}\  {\bf B 692}, 36 (2010)
  \texttt{e-Print arXiv:1003.1041 [hep-ph]}.
  %%CITATION = PHLTA,B692,36;%%

%\cite{Thompson:2008qw}
\bibitem{Son-Thompson}
  E.~G.~Thompson and D.~T.~Son,
  %``Magnetized baryonic matter in holographic QCD,''
  Phys.\ Rev.\  D {\bf 78}, 066007 (2008)
  [arXiv:0806.0367 [hep-th]].
  %%CITATION = PHRVA,D78,066007;%%


\bibitem{Kolb}
Peter F. Kolb, Ulrich W. Heinz, SUNY-NTG-03-06, Invited review for
{\it ``Quark Gluon Plasma 3''}. Editors: R.C. Hwa and X.N. Wang,
World Scientific, Singapore, 634-714 (2003).



\bibitem{Andersen}
Jens O. Andersen, Michael Strickland, Nan Su, ``Three-loop HTL
gluon thermodynamics at intermediate coupling'', {\it JHEP} {\bf
1008}, 113, 2010; \texttt{e-Print arXiv: 1005.1603 [hep-ph]}.


\bibitem{Fukushima:2009ft}
  K.~Fukushima, D.~E.~Kharzeev and H.~J.~Warringa,
  ``Electric-current Susceptibility and the Chiral Magnetic Effect,''
  Nucl.\ Phys.\  A {\bf 836}, 311 (2010);
  \texttt{e-Print arXiv: 0912.2961 [hep-ph]}.
  %%CITATION = NUPHA,A836,311;%%


\bibitem{son}
  D.~T.~Son and M.~A.~Stephanov,
  ``Axial anomaly and magnetism of nuclear and quark matter,''
  Phys.\ Rev.\  D {\bf 77}, 014021 (2008);
  \texttt{e-Print arXiv: 0710.1084 [hep-ph]}.
  %%CITATION = PHRVA,D77,014021;%%


\bibitem{gv}
  A.~Gorsky and M.~B.~Voloshin,
  ``Remarks on Decay of Defects with Internal Degrees of Freedom,''
  Phys.\ Rev.\  D {\bf 82}, 086008 (2010);
  \texttt{e-Print arXiv: 1006.5423 [hep-th]}.
  %%CITATION = PHRVA,D82,086008;%%
\end{thebibliography}
